\documentclass[12pt]{article}
\pdfoutput=1
\usepackage[T1]{fontenc}
\usepackage{fullpage}
\usepackage{amsmath,amsfonts,amssymb,graphicx,color}

\usepackage{authblk}

\usepackage{amsthm}
\usepackage{subcaption}

\usepackage{listings}
\usepackage{mathrsfs}

\usepackage[unicode=true,bookmarks=true,
bookmarksnumbered=false,bookmarksopen=false,
breaklinks=false,
pdfborder={0 0 1},
backref=false,colorlinks=true]{hyperref}
\usepackage{xcolor}
\usepackage{verbatim}
\usepackage{cite}
\usepackage[hang,flushmargin]{footmisc}

\hypersetup{pdftitle={Wormhole Traversability via Quantum Random Walks},
 pdfauthor={Ning Bao},
 citecolor=black,linkcolor=black,urlcolor=black}

\newcommand{\ket}[1]{\left| #1 \right\rangle}
\newcommand{\bra}[1]{\left\langle #1 \right |}
\def\({\left(}
\def\){\right)}
\def\[{\left[}
\def\]{\right]}

\newcommand{\be}{\begin{equation}}
\newcommand{\ee}{\end{equation}}

\newcommand{\Fig}[1]{Fig.~\ref{#1}}

\newtheorem{thm}{Theorem}[section]

\newtheorem{example}[thm]{Example}

\title{Wormhole Traversability via Quantum Random Walks}
\author[1,2]{Ning Bao}
\author[1]{Vincent P. Su}
\author[1]{Mykhaylo Usatyuk}
\affil[1]{Berkeley Center for Theoretical Physics, Berkeley, CA, 94720, USA}
\affil[2]{Computational Science Initiative, Brookhaven National Lab, Upton, NY, 11973, USA}

\date{\today}

\begin{document}

\maketitle

\begin{abstract}
In this work we seek to generalize the connection between traversable wormholes and quantum channels. We do this via a connection between traversable wormholes and graph geometries on which we can perform quantum random walks for signal transmission. As a proof of principle, we give an example of a geometry for which quantum random walk signal traversal occurs exponentially faster than a standard ballistic classical traversal. We then generalize this connection to superpositions over holographic states, and argue that the quantum random walk/graph geometry picture works to study quantum channel properties of such states.
\end{abstract}

\section{Introduction}
A fascinating recent insight into the AdS/CFT correspondence \cite{Maldacena:1997re,Witten:1998qj} is that there is a holographic description for spacetimes with traversable wormholes of the type described by \cite{Morris:1988tu} via double trace deformations of the boundary theory \cite{Gao:2016bin,Maldacena:2017axo}. These objects can be understood in the language of quantum channels as bipartite Hamiltonian-type quantum communication channels \cite{Bennett:2003} that allow for transmission of quantum information between conformal boundaries \cite{Bao:2018msr}. While the boundary details of the quantum channel are somewhat opaque, from the perspective of the bulk theory the transmission of information is simply the motion of a particle through a classical gravity theory from one conformal boundary to the other enabled by the negative energy shock-wave dual to the boundary double trace deformation as described in \cite{Gao:2016bin,Maldacena:2017axo}.

While the traversable wormhole story is a compelling one, it nevertheless possessed some limitations in its initial formulation. One of these, the non-eternal nature of the traversable wormhole solution, was addressed by \cite{Maldacena:2018lmt}. The other limitation is the requirement of large energy per qubit in order to guarantee sufficiently small spatial support for the qubit to ``fit through'' the wormhole.

Indeed, we wish to go to the opposite limit, that of signals with very large spatial support, in order to study possible interference phenomena of topologically nontrivial wormhole configurations. We will argue that such interference phenomena are quite reminiscent of the double-slit experiment,
 which has been shown to be equivalent to quantum
random walks \cite{tang2018experimental}, another active area of research in quantum information
science, as reviewed in \cite{kempe2003quantum, childs2009universal}. Given this equivalence, we will argue that there exist spacetime geometries for which normal ballistic particle propagation in the bulk is significantly slower in transferring quantum information than a channel based on quantum interference effects. We will cover in detail an example of a network of traversable wormholes for which quantum random walk behavior dramatically outperforms their classical counterpart.

The organization of this work will be as follows. In section 2, we will give minimal but sufficient overviews of the necessary details of the quantum random walk and traversable wormhole stories. In section 3, we will argue that sending a certain type of holographic excitation into the bulk corresponds to implementing the quantum random walk procedure, and we will give an example of a spacetime geometry for which such a procedure allows for much faster rates of quantum information transfer than simple ballistic protocols from the perspective of the bulk of the form considered by \cite{Gao:2016bin, Maldacena:2017axo}. In section 4, we will further extend this paradigm by considering quantum walk behavior, not on graphs corresponding to a single classical bulk geometry described by general relativity, but rather on graphs corresponding to superpositions of classical spacetime geometries, in extension of the cases considered by \cite{almheiri2017linearity}. Here, we find that quantum superposition can lead to nonzero quantum channel capacity even in situations where the dominant saddle in the superposition has zero capacity. Finally, we offer some concluding comments in section 5.

\section{Background}
\subsection{Quantum Random Walks}

We begin by introducing the notion of a classical random walk. Consider a classical graph $G$, defined by a fixed set of vertices connected by a set of undirected edges of given edge weights. The adjacency matrix $A(G)$ of this graph is defined by labelling the rows and columns by the vertices given, and having the $i^{th}$ row and $j^{th}$ column of the adjacency matrix contain the edge weight of the edge connecting vertex $i$ and vertex $j$, with a $0$ inserted if the two edges are not connected. A length $k$ random walk is a collection of vertices $v_1$ through $v_k$ for which every $v_i, v_{i+1}$ are connected in $G$. Powers of the adjacency matrix $A^k$ can be used to collect statistics on the random walk.

A continuous time quantum random walk, as reviewed in \cite{kempe2003quantum,childs2009universal}, is a quantum generalization of a classical random walk on vertices. In the quantum case, vertices are replaced by state vectors, and the dynamics are governed by a Hamiltonian whose matrix elements are related to the adjacency matrix $A$. Since time is no longer discretized, we introduce a factor $\gamma$ to describe the rate at which transitions between states happen. For weighted graphs, $\gamma$ will not be uniform. The adjacency matrix $A$ has a continuous generalization $M_{ab}$.

\begin{equation}
M_{ab}(G) =
  \begin{cases}
    - \gamma(a,b) & a \neq b, \text{$a$ shares an edge with $b$} \\
    0 & a \neq b, \text{$a$ does not share an edge with $b$} \\
    \sum_{(a,c)} \gamma(a,c)  & a = b, \text{$a$ and $c$ share an edge}
  \end{cases}
\end{equation}

The system is then described by typical Schrodinger evolution.
\be
\ket{\psi(t)}=e^{iM_{ab}(G)t}\ket{\psi_0}.
\ee

In addition to its universality for quantum computation \cite{childs2009universal} and its well-suitedness to spatial search\cite{childs2004spatial}, quantum random walks can also be used for rapidly traversing a graph from a labeled ``in'' point to a labeled ``out'' point. Indeed, there exist example graphs for which the quantum random walk not only dramatically outperforms the classical random walk for graph traversal, but has an exponential speedup over any known classical algorithm; see \cite{kempe2003quantum} for details.

\begin{figure}
\begin{subfigure}{.5\textwidth}
  \centering
  \includegraphics[width=.9\linewidth]{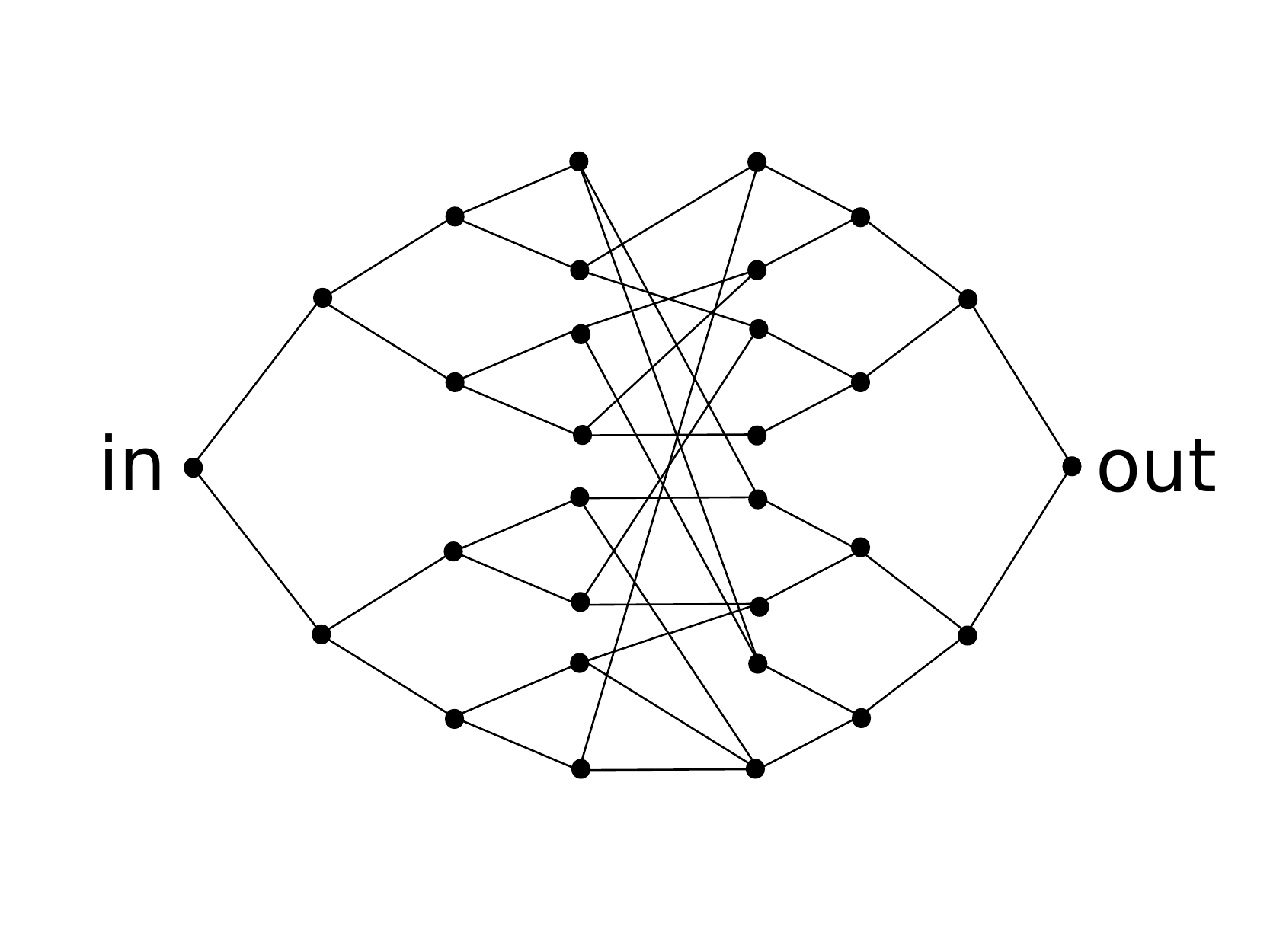}
  \caption{}
  \label{fig:glued_trees}
\end{subfigure}%
\begin{subfigure}{.5\textwidth}
  \centering
  \includegraphics[width=.8\linewidth]{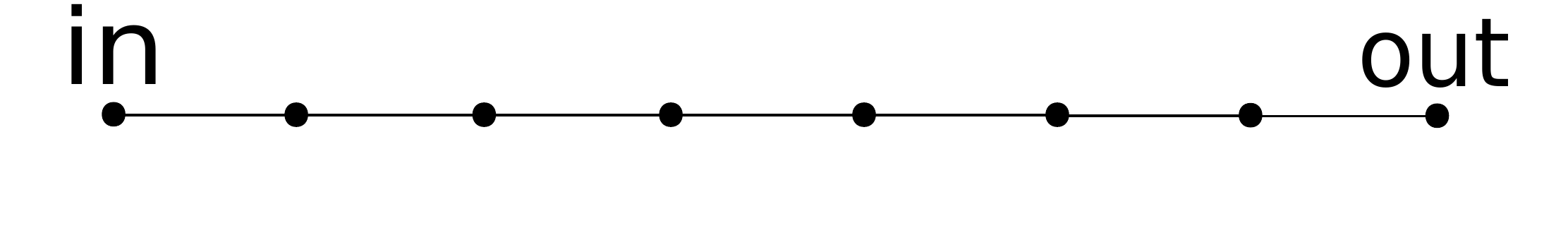}
  \caption{}
  \label{fig:linear_walk}
\end{subfigure}%
\caption{a) A glued trees graph consists of two binary trees, each of depth $n$. The leaves of each tree are then connected to two leaves of the other tree. All edges have uniform weight. Classical algorithms take exponential time in $n$ to traverse from the left root to the right. For the quantum walk evolution, the state vectors correspond to nodes. Additionally, the quantum random walk allows for the state to be in superpositions of nodes, allowing for quantum effects such as interference. b) The equivalent graph for a quantum random walk on the glued trees graph. Because of symmetry in the graph, the state space can be reduced from the set of all nodes to the set of column indices. More explicitly, let $\ket{j} = \sum_{i \in \text{col }j}\ket{i}$. The set of states $\ket{j}$ are closed under the quantum random walk evolution specified by the glues treed graph. This allows the original graph to be cast as a quantum random walk on a line.}
\label{fig:graphs}
\end{figure}

An example of such a graph is the glued trees graph, in which two binary trees of depth $n$ and constant edge weight are ``glued together'' at their leaves. The gluing procedure connects each leaf on the left tree to exactly two leaves on the right tree at random, and vice versa. See \Fig{fig:graphs}. The intuitive explanation for the struggle of the classical algorithms to traverse such a graph are the glued regions; a local classical search algorithm cannot differentiate between making progress in that region or being rerouted back towards its starting point. Therefore, a bottleneck exponential in the depth of the trees arises as the algorithm must generically exhaustively check a large fraction of the leaves \cite{aaronson2018forrelation} of both trees. By contrast, it has been shown that a quantum random walk algorithm can traverse this graph in a time polynomial in the depth \cite{kempe2003quantum,childs2004spatial}.

Generically, quantum random walks only provide a quadratic speedup over their classical counterparts. However, we briefly describe the argument explained in \cite{childs2002example}, that shows the exponential speedup of quantum random walks for these glued tree graphs. First, the subspace of states is significantly reduced because of the symmetry of the Hamiltonian (as a consequence of the graph structure). Consider the states which are defined as superpositions of all vertices within a given column. Because $H$ preserves states within this subspace, the problem is equivalent to a random walk along a line. In the limit of an infinite line, one can calculate the exact speed of propagation for this example, which is $2\sqrt{2} \gamma$, implying that the hitting time (the time at which $\bra{\psi_f}e^{iHt}\ket{\psi_i}$ first becomes nonzero) is linear in $n$. We note that for this scenario, the quantum random walk outperforms the classical one not in reaching the middle of the graph, but in progressing away from the exponentially large middle region.

\subsection{Traversable Wormholes: the Bare Basics}
There are many technical details involved in the traversable wormhole construction, but as many of them will not be relevant for the discussion at hand, we will only include the necessary ingredients. For the reader interested in more technical discussions, see \cite{Gao:2016bin,Maldacena:2017axo,Bao:2018msr}. For simplicity, our description here will correspond to the description of \cite{Gao:2016bin}, as opposed to those of later works, for example \cite{Maldacena:2017axo,Maldacena:2018lmt}.

Normal wormholes in general relativity are not traversable, as the Null Energy Condition prevents wormholes from being traversable. More specifically, any positive energy signal that is sent into one of the wormhole mouths will fail to get through, as it triggers the pinching off of the wormhole. This pinching prevents the signal from exiting out of the other wormhole mouth \cite{Morris:1988tu}, and therefore also preserves causality. The only way to prevent this from occurring is by stabilizing the wormhole with a source of negative stress-energy, to prevent the wormhole from collapsing. While such stabilization is normally considered unphysical in general relativity, it was first pointed out in \cite{Gao:2016bin} that, in the context of the AdS/CFT correspondence, such negative stress energy corresponds to a noncontroversial double trace deformation from the perspective of the boundary conformal field theory. First, begin with a thermofield double state corresponding to a marginally nontraversable wormhole \cite{maldacena2003eternal}, of the form

\be
\ket{\Psi} = \frac{1}{\sqrt{Z}} \sum_n e^{-\beta E_n/2} \ket{\bar{n}}_L \otimes \ket{n}_R.
\ee

A negative energy shock-wave \cite{Shenker:2013pqa} can now be inserted on both conformal boundaries by the activation of a double trace deformation, of the form
\be
\delta S = \int {\rm d}t \,{\rm d}x\, h(t,x)\mathcal{O}_R(t,x)\mathcal{O}_L(-t,x),\label{eq:dbltraceaction}
\ee
where $\mathcal O$ has conformal dimension $\Delta$ and the coupling $h$ has support only in some time window. It is worth re-emphasizing that this deformation is totally innocuous from the perspective of the boundary conformal field theory.

\begin{figure}
\centering
\includegraphics[width = 5cm]{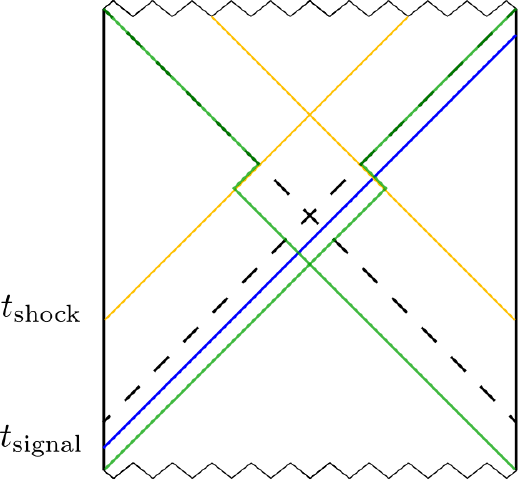}
\caption{Penrose diagram for the traversable AdS wormhole. We start with a two sided BTZ black hole \cite{Banados:1992wn} and send in a signal at time $t_{\text{signal}}$ (blue). After at least a scrambling time, we turn on the double trace deformation (yellow) at time $t_{\text{shock}}$ whose negative energy has the effect of moving the apparent horizon (green) to the final event horizon (black dashes). Without the double trace deformation the signal cannot propagate between the conformal boundaries. Hence, the wormhole has been rendered traversable.}
\label{fig:geometry}
\end{figure}

The negative energy shock shifts the horizon location to a smaller radius. This has the effect of opening up a traversable path through the wormhole for null particles sent significantly before the boundary double trace deformation is applied. This procedure thus allows particles to propagate between the two conformal boundaries. For a summary, see \Fig{fig:geometry}. 

Traversable wormholes of this type have a bound on the minimum amount of energy per qubit that can be sent through, as discussed in \cite{Gao:2016bin, Maldacena:2017axo}. Indeed, this restriction was a key ingredient in the discussion of the quantum channel capacities of that class of traversable wormholes done in \cite{Bao:2015nqa}, which related the traversable wormholes to bipartite Hamiltonian quantum channels as discussed by \cite{Bennett:2003}. This bound is essentially to ensure that each qubit fits through the open portion of the wormhole throat, instead of being ``too large to fit.'' This restriction essentially obviated the necessity to consider interference effects. This did not need to be the case, however, something which will be the focus of the latter portions of this paper.

For the remainder of this work, we will proceed under the assumption that the deformation necessary to engineer wormhole traversability is something that can be performed modularly. Namely, in spacetime geometries containing many wormholes, any subset of these wormholes (in particular, all of them) can be rendered traversable on the worldline of a particle attempting a traversal from one conformal boundary to the other\footnote{In fact, this problem can be obviated when considering eternal traversable wormholes, of the type studied by \cite{Maldacena:2018lmt}}. We view this as a problem of tuning the double trace deformation, and take the existence of a double trace deformation that renders a single wormhole traversable as a theoretically sufficient proof of principle. One way of potentially doing so is by invoking the conjectured ER/EPR correspondence in the limit of classical wormholes \cite{Maldacena:2013xja}, which has been tested for consistency in this limit \cite{Bao:2015nqa, Bao:2015nca,Remmen:2016wax}. Such a construction would allow one to consider multiple wormholes in the same bulk geometry, as at sufficiently small scales asymptotically AdS space functions very similarly to asymptotically flat space.

\subsection{Entanglement-Assisted Quantum Channels}

We briefly review the notion of entanglement-assisted quantum channels. For a more detailed overview, with a focus on the connection with traversable wormholes, see \cite{Bao:2018msr}. A quantum channel is defined as a completely positive trace preserving map on the space of density matrices. A pedestrian caricature of a channel is a (possibly noisy) process that transmits a quantum state that Alice possesses to her friend Bob. The channel capacity is related to how faithfully the channel transmits information, in other words, whether the channel can be inverted to recover the initial density matrix. This is distinct from the notion of channel rate, which is the physical speed with which the channel transmits information. For an entanglement-assisted channel, one considers the same process of sending a quantum state from Alice to Bob, under the condition that they share some number of entangled bits. In some cases, entanglement as a resource can dramatically boost the channel capacity \cite{Bennett:1999hf}, or even create channel capacity for which none would exist in the absence of entanglement. Indeed, for the basic quantum teleportation protocol \cite{Bennett:1993:teleportation,Nielsen:2011:QCQ:1972505}, Alice and Bob need to start with a pair of entangled bits that gets consumed in the process. Typically, entanglement is treated as a resource which needs to be generated, but from here we assume that entanglement is free, namely that it already exists for the channels we consider.

We claim that traversable wormholes are indeed a special case of an entanglement-assisted quantum channel. The fact that a signal can propagate through the bulk between the two boundaries is a special property of the thermofield double state and is not true for generic CFT states. We argue that the entanglement structure of the thermofield double exactly corresponds to the entanglement-assistance needed to make the quantum channel capacity nonzero between Alice and Bob who are thought of as living on opposite boundary CFTs. For states in the CFT which do not have entanglement, e.g. various product states, the channel capacity will be zero because the premise of entanglement resources is not satisfied.

\section{Holographic Signals dual to Quantum Walk Traversal}
Now let us return to the holographic spacetime picture. It should be immediately noted that any space-like slice of a holographic spacetime can be discretized into a graph geometry. The regions of spacetime connected by the wormholes would correspond to the vertices, and the wormholes that connect these regions would become edges with edge weight given by the area of the throat of the wormhole. To encode distances in the graph, we can insert additional vertices so edges between vertices have unit length. Moreover, the vertices corresponding to the CFT boundaries must be connected to the remainder of the graph by at most one edge to correspond to a classical bulk geometry, to avoid problems with being non-Hausdorff as discussed in \cite{Bao:2015nqa,bao2018bulk}. Such techniques were used to great effect in the study of the holographic entropy cone \cite{bao2015holographic} and subsequent work on holographic entropy relations \cite{hubeny2018holographic,hubeny2018holographic2}.

One natural question is the bulk interpretation of a quantum random walk traversal from the vertex corresponding to a conformal boundary to the vertex corresponding to the other conformal boundary. To answer this question, one can consider a space-filling signal sent with support on the entirety of a conformal boundary and consider its propagation into the bulk. As a specific case, let's consider that this conformal boundary is connected to another conformal boundary by a geometry given by two pairs of pants stitched together at the legs, with the waists corresponding to the two separate conformal boundaries.

In this case, the effect of the geometry will be to induce the gravitational version of the double slit experiment; the bulk metric quite literally provides two slits, with recombination of the signal on the other side of the slits. It has been shown in previous work \cite{tang2018experimental} that double slit experiments (and generalizations thereof) are equivalent to evolution by a quantum random walk, and thus the propagation of such a signal would also be equivalent to a quantum random walk.

We can formally define the quantum walk traversal as follows. Let $\ket{\psi}$ denote the holographic CFT state dual to the stitched pants geometry defined above. At the appropriate time, we apply the double trace deformation or sequence of double trace deformations that render the wormholes traversable. Let $O_L$ denote an operator that creates a diffuse signal, originating on the left conformal boundary, propagating into the bulk. We can act with this operator on our state $\ket{\psi}$ and time evolve to get $U(t)O_L \ket{\psi}$. In this new state, the signal has had time $t$ to propagate through the bulk and we would like to determine the probability that the signal has reached the right conformal boundary. One way to do this is to prepare a reference state $\ket{\psi_{ref}}$ where the signal is created on the right boundary. Suppose the operator that does this is $O_R$. The correct reference state is $\ket{\psi_{ref}}=O_R U(t)\ket{\psi}$. The probability that our signal has made it through the bulk is proportional to $\bra{\psi} O^{\dag}_R(t) O_L(0) \ket{\psi}$. Note that this expectation value is evaluated with the initial state $\ket{\psi}$ held at some reference time. As we change $t$, we can probe how much of the signal makes it through the wormhole. This procedure readily generalizes to the task of sending signals with arbitrary states.

Because we are free to consider bulk geometries so long as they correspond to a graph\footnote{This is true at least in the case of three-dimensional gravity, where orbifold identifications of the boundary give us this freedom \cite{balasubramanian2014multiboundary}. The method of constructing the glued pair-of pants geometry \cite{Brill:1995jv,Brill:1998pr, Aminneborg:1997pz} in the previous section is clear, and we expect other, more involved geometries to be possible as well, modulo the concerns of \cite{marolf2017handlebody}.}, an immediate question arises. Do there exist any graphs, and thus correspondingly bulk geometries, for which the quantum random walk implementation of a quantum channel between the two CFT's outperforms that of the quantum channel dual to classical ballistic bulk motion as discussed in \cite{Gao:2016bin,Maldacena:2017axo,Bao:2018msr}? The answer to this seems to immediately be yes: let the graph considered be the glued tree graph. The classical ballistic motion on this graph is a subset of the set of all currently known classical traversal protocols, and thus is known to be slower at reaching the other conformal boundary than the quantum random walk on the same graph. Thus, by invoking the duality, it is clear that for bulk geometries equivalent to the glued trees graph connecting two conformal boundaries, the propagation suggested at the beginning of this section exploiting interference phenomena transmits quantum information from one conformal boundary to the other exponentially faster than the  classical ballistic protocol.

It should also be noted that since these signals can have lower energy than the signals considered in the original traversable wormhole setup, the wormhole could potentially remain traversable for more signals before it collapses. In such situations, our diffuse signals should be understood as plane waves as opposed to localized gaussian wavepackets designed to ``fit through'' traversable wormholes.

Also note that in the preceding sections we considered eternal time independent spacetimes; for dynamical spacetimes the graphs become time dependent which can be captured by a time dependent adjacency matrix. This area is less well understood and will be left for future work.

\section{Superposing Bulk Geometries}

Having discussed quantum random walks in the context of a single classical bulk geometry, we now turn our attention to quantum random walks over superpositions of classical bulk spacetime geometries. Via the duality, this corresponds to asking about signal propagation from one conformal boundary to another when the CFT state is a superposition of states individually  dual to classical spacetimes, but for which the superposed state does not necessarily correspond to a classical spacetime geometry. This setup motivates the question: what can we say about information transfer between conformal boundaries when the CFT state is a superposition over holographic states?

A thing to note about such superpositions is that the previous requirement that each CFT vertex is attached to the rest of the graph by at most one edge is no longer true: since the bulk geometry is no longer classical, concerns of the type invoked in \cite{Bao:2015nqa,bao2018bulk} no longer apply. This property fundamentally differentiates between graphs corresponding to classical bulk geometries and superpositions thereof.

Even with multiple edges from each CFT vertex, we can still preserve the graph
structure that is needed for quantum random walks. The weights for each edge
corresponding to subgraphs of different geometries can be taken to be the
probabilistic weighting of terms in the superposition. Excitingly, because the
set of geometric states forms an overcomplete basis for the Hilbert space, any
state of two CFTs can be decomposed into a superposition of classical bulk
geometries. From this superposition we can construct a graph and protocol for sending information from one CFT to the other! This allows for the usage of the quantum walk algorithm to diagnose transmission of quantum information from one CFT to the other for generic states.

\begin{figure}
\centering
\includegraphics[width = 15cm]{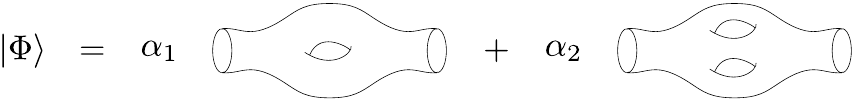}
\caption{A superposition of two states with classical bulk geometries. The superposition is of two CFT states so the figure should be taken as a heuristic description that is sufficient for the purpose of calculating correlation functions. The bulk geometries have different topologies and contain wormholes that link the two conformal boundaries. The geometries in the above figure should be understood as spacelike slices of the bulk duals to the states $\ket{\psi_{1}},\ket{\psi_{2}}$. We can act on this state with $O_L$ to send in a diffuse signal through both wormholes. The probability that our signal is detected on the other side is a weighted sum over the probabilities that the signal traverses the individual wormholes.}
\label{fig:superpositionstate}
\end{figure}

\subsection{Superposition of two Holographic States}
We first work through an instructive toy example. Consider two holographic states $\ket{\psi_1}$ and  $\ket{\psi_2}$ that are distinguished from each other by their non-trivial bulk topologies. As an example, we can take the two states to have a different number of handles in the bulk. We will also assume that these states contain a traversable wormhole that links the two conformal boundaries. We consider a superposition of these states

\begin{figure}
\centering
\includegraphics[width = 15cm]{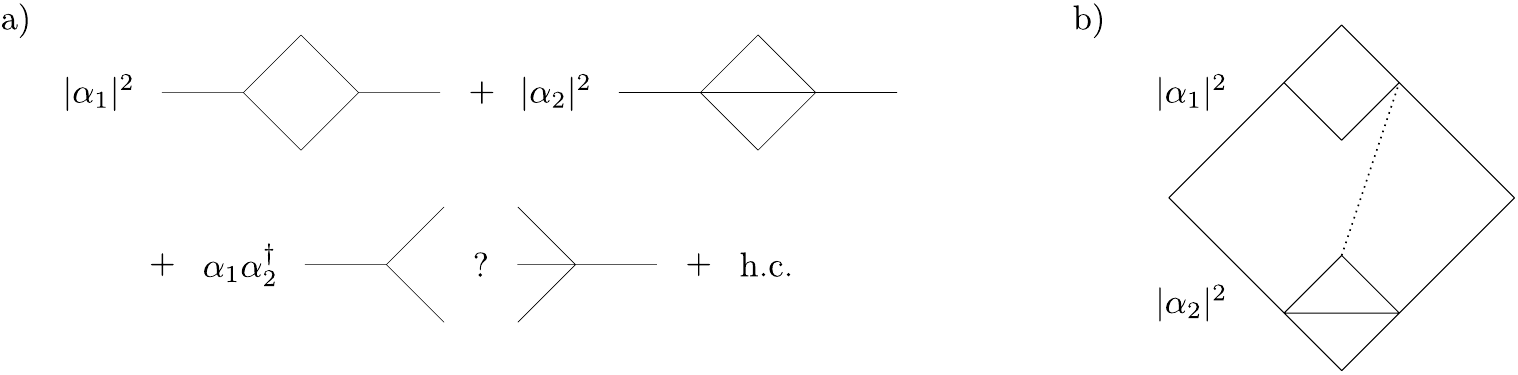}
\caption{ a) The quantum random walk interpretation should be understood as a walk over the above graphs with the given weightings. A signal is introduced on the left-most vertex and allowed to propagate through the graph. We heuristically show one cross term with coefficient $\alpha_1 \alpha_2^{\dag}$, the second cross term is the hermitian conjugate of the shown cross term. The cross term is exponentially suppressed in the central charge and can be interpreted as a tunneling between geometries. b) We can heuristically merge these graphs into one graph with one adjacency matrix. The signal propagates through both graphs simultaneously; to remain consistent with a), wavefunctions from different branches should be understood as not interfering. The dashed line in the interior of the graph should be interpreted as exponentially suppressed tunneling between the different bulk geometries.}
\label{fig:superpositiongraph}
\end{figure}

\be
\ket{\Phi}=\alpha_1 \ket{\psi_1} + \alpha_2 \ket{\psi_2}
\ee

\noindent where $\alpha_i$ are complex numbers. Generally, this new state will not be dual to a classical bulk geometry. Nonetheless, we can apply our prescription from the previous section to independently perturb the left CFT with an operator $O_L$ and determine how long it takes the information to spread to the right CFT. Introducing such a perturbation and letting the signal propagate into the bulk for a time $t$ we get the following state.

\be
U(t)O_L \ket{\Phi} = U(t)(\alpha_1 O_L \ket{\psi_1} + \alpha_2 O_L \ket{\psi_2})
\ee

We can prepare a reference state $\ket{\Phi_{ref}}$ where the initial signal is inserted on the right boundary after the unperturbed state is time evolved. Suppose the operator that accomplishes this is $O_R$. The reference state is

\be
\ket{\Phi_{ref}} = \alpha_1 O_R U(t) \ket{\psi_1} + \alpha_2 O_R U(t) \ket{\psi_2}
\ee

The probability that our signal will be detected on the right boundary is given by

\be
  \bra{\Phi} O_R^{\dag}(t)O_L(0)\ket{\Phi} \approx
  |\alpha_1|^2 \bra{\psi_1} O_R^{\dag}(t)O_L(0) \ket{\psi_1}
  + |\alpha_2|^2 \bra{\psi_2} O_R^{\dag}(t)O_L(0) \ket{\psi_2}
\ee

\noindent where we have dropped cross terms between different holographic states. In \cite{almheiri2017linearity} it was shown that the inner product between different thermal holographic states is exponentially suppressed in the central charge. We expect a similar behavior for our holographic states so we are justified in ignoring cross terms, at least for superpositions over small numbers of geometries.

The probability that our signal reaches the right CFT is a sum over weighted probabilities of the signal traversing the wormhole of each individual classical geometry. The quantum random walk interpretation of this process is shown in \Fig{fig:superpositiongraph}. The bulk geometries are discretized into graphs and a signal is inserted on the vertex of the graph corresponding to the left CFT boundary. The signal undergoes a quantum random walk through the graph and some portion of it traverses the graph. The probability that our signal arrives at the right CFT is a sum over weighted probabilities of the quantum random walk traversing each graph. The cross terms correspond to tunneling between classical graphs. This effect is exponentially suppressed in the central charge and can become important if the superposition is over $e^{O(c)}$ states; this is a subtle point and we return to it in the discussion.

\subsection{Entangled CFT States}
We can extend the above discussion to more generic CFT states without bulk duals by utilizing the previously mentioned property that holographic states form an overcomplete basis for our CFT Hilbert space. An intuitive way to understand this follows. The thermofield double state $\ket{\Psi(\beta)}$ is a sum over energy eigenstates $\ket{\bar{n}}_L \otimes \ket{n}_R$ with coefficients weighted by a Boltzmann factor. These states are linearly independent for different inverse temperatures $\beta$ and we can invert them to obtain the energy eigenstates as a sum over thermofield double states of different temperatures. To complete this argument we need to find a decomposition for states of the form $\ket{\bar{n}_i}_L \otimes \ket{n_j}_R$ where $i\neq j$. These states are tensor products of energy eigenstates of a single CFT, $\ket{n}$, which are black hole microstate geometries and holographic. Thus, every CFT state can be written as a superposition over holographic states.

When writing a generic state $\ket{\psi}$ as a sum over holographic states it is important to keep in mind that our analysis only works if our sum contains fewer than $e^{O(c)}$ terms. This is important because different decompositions of $\ket{\bar{n}}_L \otimes \ket{n}_R$ will correspond to sums over different numbers of holographic states. For example, a state that can be written as a superposition of two wormhole states can equivalently be written as a sum over a much larger number of disconnected geometries. However, the sum over disconnected geometries can be over exponentially many terms, and the assumption that off-diagonal terms are suppressed breaks down. We will further address the exponential upper bound in the discussion section.



Using the above prescription we can write any state $\ket{\psi}$ as a superposition over holographic states. If the state $\ket{\psi}$ has nontrivial entanglement structure then the superposition will contain terms dual to two sided eternal black holes with different throat sizes. Turning on a double trace deformation or sequence of trace deformations, as discussed previously, will render these geometries traversable. From the graph perspective, a signal simultaneously propagates through all graphs in the superposition. Most of these graphs are disconnected, but the graphs corresponding to traversable wormholes provide a pathway for the signal to propagate to the second boundary. A state $\ket{\psi}$ with more entanglement of the form $\ket{\bar{n}}_L \otimes \ket{n}_R$ will have a larger contribution from connected graphs, so the quantum channel capacity of our state scales with the entanglement of the state. One interesting thing to note is that if the entanglement of our state is between low energy eigenstates then the corresponding graphs are dual to disconnected copies of thermal AdS. To leading order, such graphs have no channel capacity, which implies that bulk quantum channel capacity arises primarily from entanglement between high energy degrees of freedom. In particular, the capacity should be set only by the total number of signals that can be sent through those wormholes that have had traversable paths opened within.

\subsection{Hawking-Page Transition}
So far our analysis has focused on superpositions of holographic states, but we can equivalently consider a situation where one state is dual to a superposition of classical geometries. This is precisely the situation described by the Hawking-Page phase transition\cite{hawking1983thermodynamics,witten1998anti} where there are two classical bulk solutions that compete for saddle-point dominance.

The thermofield double state $\ket{\Psi(\beta)}$ is dual to the eternal AdS black hole above a critical temperature $\beta < \beta_{crit}$. Below the critical temperature the geometric dual is two copies of thermal AdS. We will work in the intermediate regime where neither saddle dominates and the state can be understood as a superposition of both classical geometries. In the large $c$ limit, this regime is sharply peaked about $\beta = \beta_{crit}$.

\be
    \ket{\Psi(\beta)} \sim \alpha_{1} \ket{BH} + \alpha_{2} \ket{AdS}
\ee

The above equation should be interpreted as a heuristic map between geometric duals and the CFT state. The weighting of a classical geometry in the superposition should be proportional to $e^{-S_E}$ where the euclidean action is evaluated for the bulk metric.

We turn on a double trace deformation which makes the eternal black hole wormhole traversable. The two copies of thermal AdS are disconnected, so the double trace deformation does not create a traversable bulk in this geometry. When we perturb the left CFT with an operator $O_L$, our signal can reach the right conformal boundary by traversing the wormhole in the black hole geometry. In the quantum walk interpretation, the signal would simultaneously propagate through a disconnected graph, corresponding to two copies of thermal AdS, and through a connected graph corresponding to the traversable black hole geometry. A portion of the signal will make it through the black hole graph and reach the second boundary. In this example, the superposition of two classical geometries plays a vital role in supporting bulk channel capacity between the two conformal boundaries.

We can also consider the regime where the thermal AdS solution dominates. In analyzing the signal propagation from the graph perspective we should include the exponentially suppressed black hole graph alongside the thermal AdS graph. In this case the black hole geometry contributes to the channel capacity despite not being the dominant saddle. From the perspective of the thermal AdS geometry it will look like the signal tunneled between two disconnected spacetimes.

It is clear that the dominant contribution to the information channel capacity does not need to come from the dominant saddle, should one exist. In the above example, the dominant saddle is two copies of thermal AdS, for which there is no bulk path connecting the two boundaries; to leading order, this bulk geometry does not contribute to the channel capacity. However, the subdominant saddle consisting of a connected wormhole will result in some channel capacity, and thus will allow for bulk transmission of information between the two boundaries even when the dominant saddle is disconnected.

\section{Discussion}
We briefly summarize our results before addressing further subtleties that may be interesting for future work.

Here we consider a new way of sending signals that are diffuse and low energy through traversable wormholes. Because the traversable wormhole is stabilized by negative stress-energy shockwaves, signals with high energy density would collapse the wormhole sooner than lower energy signals, thus allowing fewer signals through the wormhole. By exploiting graph discretization of the bulk geometry, we then leverage the connection between quantum signal propagation and quantum random walks to analyze traversable wormholes as entanglement assisted quantum channels. Specifically, it allows us to construct a procedure for analyzing the channel rate for arbitrary wormhole geometries, and even to construct an example geometry for which this manner of interference-driven wormhole traversal is exponentially faster than a classical ballistic traversal of the same geometry.

Furthermore, we exploit the fact that generic CFT states not dual to a classical bulk geometry can be decomposed into superpositions of such states. This allows us to extend our procedure to a new regime of states that usually evade holographic analysis. By linearity of quantum mechanics, we can consider the graph procedure on each of the individual states. As long as the number of states in the superposition is bounded above by an envelope function that scales as an exponential in the central charge, the channel capacity of a superposition of states can then be related to the channel capacity of individual classical states, weighted by their relative probabilities. In the next subsection we also comment on when we expect to lose computational control in this approximation.

\subsection{Unentangled States}
It is worth discussing the edge case of attempting to apply this method to
unentangled CFT states. In this case, one expects that turning on a double trace deformation does not create channel capacity. However, this would appear to be in tension with the story being told here about CFT states being described by graphs.

The resolution to this comes from the fact that unentangled states are decomposed into superpositions over product state geometries (e.g. geometries which are not connected by a bulk path). In this case the channel quantities would trivially be zero, and using a disconnected graph to compute hitting times between disconnected edges gives a trivial result.

The second point is that the double trace deformation itself could, in principle, entangle the two CFTs at subleading order. This was correctly not taken into consideration in the first pass of traversable wormhole work, as that work focused on a single geometry where the vast majority of the e-bits are unaffected by the subleading correction caused by the double trace deformation. However, in a case where the leading order number of e-bits is zero, the double trace deformation could potentially lead to a nonzero, though small, number of e-bits that could function as an entanglement resource for an entanglement assisted quantum channel. It is worth stressing that this will result in a very small number of uses of the channel before it has consumed all of its entanglement resource, but it is nevertheless a potentially non-zero effect. We will study this possibility in later work.

One final important case to consider is when a state is equivalent to a sum over $e^{O(c)}$ holographic states. As we saw in the preceding section, an entangled state of the form $\ket{\bar{n}}_L \otimes \ket{n}_R$ can be decomposed into a superposition over high temperature TFD states and equivalently as a superposition over low temperature TFD states. From the graph perspective it looks like the high temperature decomposition is over connected graphs while the low temperature decomposition is over disconnected graphs, leading to contradictory interpretations regarding the channel capacity and related channel properties. The resolution to this comes from the following: While it is true that the overlap between different holographic states is exponentially suppressed in the central charge, if the state is decomposed into an exponential in central charge number of terms the off diagonal terms will then contribute at the same order as the diagonal terms. In this case the graph analysis becomes more involved, as the off-diagonal terms would correspond to new edges in your graph/adjacency matrix, thus changing the time evolution associated with the quantum random walk.

\subsection{Path Integrals}
It is interesting to understand the picture being presented here in the context of the path integral formulation of AdS/CFT \cite{caputa2017liouville, takayanagi2018holographic}. Here, different edges coming out of the conformal boundary correspond to different weightings of the different geometries in the path integral. The graph picture appears to give a calculable implementation of the path integral proposal via the utilization of quantum random walks. As discussed in the previous subsection, there could be a loss of computational control in cases where the number of terms in the superposition becomes too large, as the off-diagonal pieces will no longer be suppressed by 1/N relative to the diagonal pieces. This can easily be seen by considering that the off-diagonal pieces are exponentially suppressed by the central charge relative to the sum of all of the diagonal terms (or one, for density matrices), and so if there is an exponential in the central charge number of nonzero diagonal terms in the density matrix, the relative suppression between the diagonal terms and the off-diagonal term goes away. This has also been discussed in \cite{Bao:2019wcf}.

\subsection{Extension to Holographic Tensor Networks}

It is also natural to consider potential extensions of this work relating quantum random walks, quantum channel properties, and graphical representations of the bulk spacetime to tensor network models of gravity such as the ones discussed in \cite{swingle2012entanglement,pastawski2015holographic,qi2013exact, hayden2016holographic,bao2019holographic, bao2018beyond,kohler2018complete, bao2015consistency}. In this case, the channel properties will describe channels related to sending signals from one portion of the conformal boundary to another portion, possibly without need of a second, boundary-disconnected conformal boundary. The tensor networks naturally provide a graph structure on which to define the quantum random walk, and so could naturally fit within this framework. Moreover, in \cite{bao2018beyond} there was also discussion of removing ``cobwebs'' in \cite{bao2019holographic} from the construction in favor of a superposition, and that superposition could potentially be treated in a similar way to what has been discussed in this work. However, there could exist subtleties in that the tensor networks do not necessarily describe all possible orientations that one can send the signal in the bulk. Therefore, we would have to construct our quantum random walk implementation in such as way as to avoid unaccounted-for interference effects from the discrete breaking of bulk isotropy by the tesselation inherent in the tensor network. We will leave this extension, and also potential connections to ideas from holographic quantum error correction \cite{almheiri2015bulk,dong2016reconstruction, cotler2017entanglement,hayden2018learning,akers2019holographic} to future work.

\section*{Acknowledgements}

We thank Scott Aaronson, Charles Cao, Aidan Chatwin-Davies, Andrew Childs, Elizabeth Crosson, Jason Pollack, Xiaoliang Qi, Grant Remmen, and Jon Sorce for useful and interesting conversations while this work was being completed; we are especially grateful to Elizabeth Crosson for early collaboration on this work.
NB is supported by the National Science Foundation under grant number 82248-13067-44-PHPXH, by the Department of Energy under grant number DE-SC0019380, and by New York State Urban Development Corporation - Empire State Development - contract no. AA289. VPS and MU gratefully acknowledge support by the NSF GRFP. This material is based upon work supported by the National Science Foundation Graduate Research Fellowship Program under Grant No. DGE 1752814. Any opinions, findings, and conclusions or recommendations expressed in this material are those of the author(s) and do not necessarily reflect the views of the National Science Foundation.
\bibliographystyle{unsrt}
\bibliography{wormholes-arXiv_v1}
\end{document}